# PHELEX: Three Series Of Measurements in the Search for Dark Photons

A. V. Kopylov[1)*], I. V. Orekhov[1)], V. V. Petukhov[1)], A. E. Solomatin[1)]

*Institute for Nuclear Research of the Russian Academy of Sciences, Moscow, Russia*

This paper presents the results of measurements of diurnal variations in the single electron count rate arising from dark photon conversion at the cathode of a gas proportional counter. Three experimental series were conducted, each consisting of 4 runs lasting approximately 60 days per run. One can expect an excess in the single electron count rate above the average level if the polarization vector of dark photons in the region traversed by the Sun forms an angle between 20° and 50° with the Earth's rotational axis. Such an effect was observed in the first series (significance 6σ) and the second series (significance 4σ). However, in the third series, which employed two identical detectors simultaneously, the effect was not observed on either detector. This result practically excludes the possibility that the effect observed earlier was purely instrumental. The results of the three series may be interpreted as a possible effect arising from the Sun's passage through regions of dark matter characterized by different orientations of the dark photon polarization vector.

## 1. INTRODUCTION

The nature of dark matter (DM) remains one of the fundamental unsolved problems in modern physics. Recent observations of gravitational lensing in a distant galaxy [1] revealed a compact DM core (size ~ kpc), which contrasts with previous data and may indicate the presence of self-interactions in dark matter. This imposes stringent constraints on possible candidates: they must possess long-range interaction and be sufficiently light to ensure high number density.

Dark photons (DP) are promising candidates satisfying both requirements: (a) they mediate long-range electromagnetic interaction (in the dark sector), (b) their assumed mass (10–40 eV) ensures a high number density in the vicinity of the Sun ($\sim 10^7$–$4\times 10^7$ cm$^{-3}$ given a local DM density of 0.4 GeV/cm$^3$ [2]). Unlike massless standard photons, DPs possess a rest mass. This leads to important consequences:

1. The Lorentz gauge condition is violated:

$$\text{div } \mathbf{A} + \partial \Phi / \partial t \neq 0, \qquad (1)$$

where $\mathbf{A}$ and $\Phi$ are the vector and scalar potentials.

2. Condition (1) is possible if

$$\mathbf{H} \neq \text{rot } \mathbf{A}, \qquad (2)$$

from which it automatically follows that

$$\text{div } \mathbf{H} \neq 0, \qquad (3)$$

$$\text{div } \mathbf{E} \neq 0, \qquad (4)$$

where $\mathbf{H}$ and $\mathbf{E}$ are the magnetic and electric field strengths, respectively.

Equations (3) and (4) mean the existence of longitudinal components in the electromagnetic field of DPs. This allows for the introduction of a polarization vector $\mathbf{P}$,

---

[1)] Institute for Nuclear Research of the Russian Academy of Sciences, Moscow, Russia
[*] E-mail: kopylov@inr.ru

associated with the direction of the **E** vector. A hypothesis arises that DM may consist of regions (domains) with a specific spatial orientation of the polarization vector. The sizes, distribution, and temporal dynamics of such domains ("granulation") are subjects of experimental research.

This hypothesis opens the possibility for a directed search. If a detector possesses anisotropic sensitivity depending on the angle between the DP's **E** vector and the detector's preferred axis, then the rotation of the Earth should lead to modulation of the detection rate – diurnal variations in sidereal time.

The PHELEX (PHoton-ELectron-EXperiment) [3–5] aims to search for DPs by detecting single electrons emitted from the surface of a metallic cathode of a gas proportional counter, presumably due to the photoelectric effect induced by DPs. In a cylindrical counter with a polished cathode, the probability of the photoelectric effect is maximum when the **E** vector is perpendicular to the counter's axis and is zero when **E** is parallel to the axis. Calculations [3] for a counter oriented along the meridian at the latitude of Moscow predict a 4-hour excess of the count rate above the average level in sidereal time, if the angle between the polarization vector of the local DM region and the Earth's rotation axis is 20°–50°. Since the orientation of the polarization vector in domains along the Sun's galactic trajectory (its orbital speed ~230 km/s) is a priori unknown, the effect is expected to be observed episodically. For reliable signal identification and background suppression, it is advisable to use a coincidence system of several detectors.

In the first two measurement series of PHELEX (four runs each, ~50–60 days/run), statistically significant excesses in the count rate ($6\sigma$ and $4\sigma$) were observed in 4-hour intervals of sidereal time (8:00–12:00 ST and 18:00–22:00 ST, respectively) [3, 4]. A key argument in favor of a possible extraterrestrial origin of the effect was its absence in solar time, which ruled out a connection with terrestrial anthropogenic processes or Solar activity. The difference in the timing of the excess between the series was interpreted as evidence of the Sun passing through DM domains with different orientations of the polarization vector. The non-uniformity of the effect's magnitude within a series also pointed to a possible change in the domain's properties along the trajectory.

The aim of the present work is to verify the results of the first series using an expanded setup, including new detectors, to rule out instrumental effects and accumulate an evidence base.

## 2. EXPERIMENTAL SETUP AND METHODOLOGY

The main PHELEX detector is a cylindrical proportional counter with a metallic cathode, whose design is detailed in [5] (see also Figs. 1, 2). Single electrons emitted from the cathode are registered.

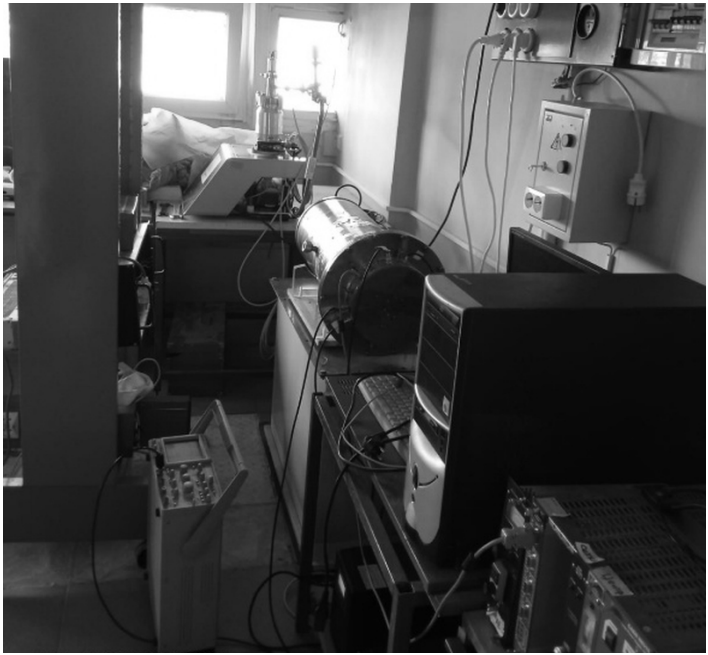

Fig. 1. Photograph of the PHELEX detector during testing.

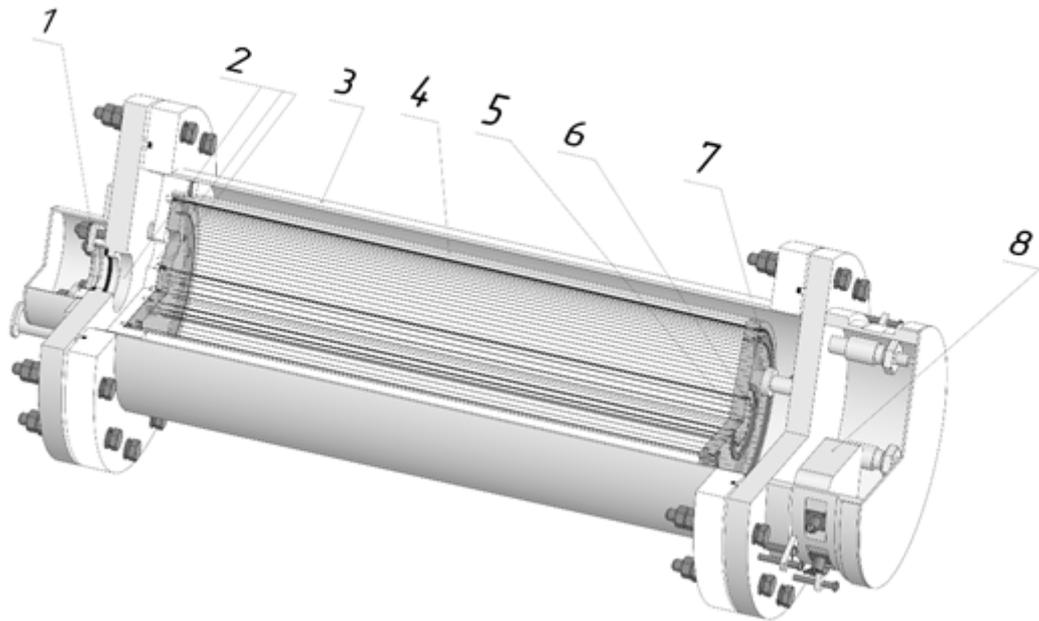

Fig. 2. Schematic of the PHELEX multicathode proportional counter design: *1* - quartz glass for ultraviolet calibration, *2* - window, *3* - steel casing, *4* - metal cathode, *5* - anode wire, *6* - first cathode wires, *7* - second cathode wires, *8* – chargesensitive preamplifier.

To test the hypothesis and rule out systematic effects, two additional detectors were developed, manufactured, and put into operation:

1. Detector Al: similar to the main detector, but with a cathode made of duralumin (a non-magnetic material). The goal is to check the dependence of the effect on the magnetic properties of the cathode (the main Fe detector has an iron cathode). The DP signal is expected to be observed on both cathode types.

2. Detector Block: operates in a mode where the electric field effectively blocks the emission of photoelectrons from the cathode, preventing their detection. Serves as a control: the signal from DPs (if it is due to the photoelectric effect) should not be observed on this detector.

### 3. RESULTS AND DISCUSSION

From late October 2024 to late June 2025, a third measurement series was conducted, consisting of four runs with a total duration of ~240 days. The active detectors registering single electrons were the main detector and the new Al detector. The Block detector was used for background control.

Data on the count rate for each run are presented as a function of sidereal time (Fig. 3 for Fe, Fig. 4 for Al) and solar time (Fig. 5 for Fe, Fig. 6 for Al).

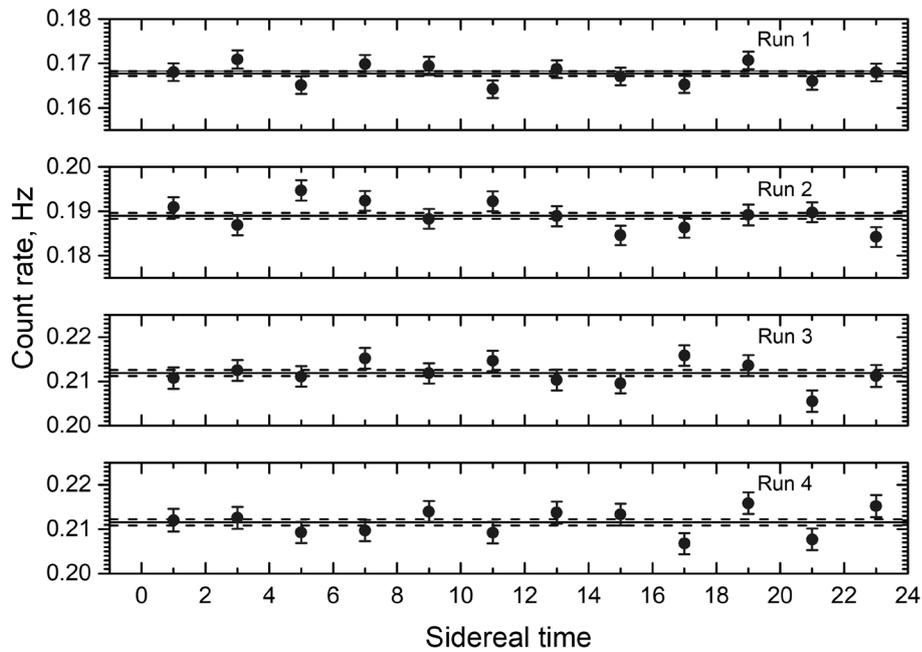

Fig. 3. Count rate of the detector with a Fe-cathode (third series) in sidereal time (four runs).

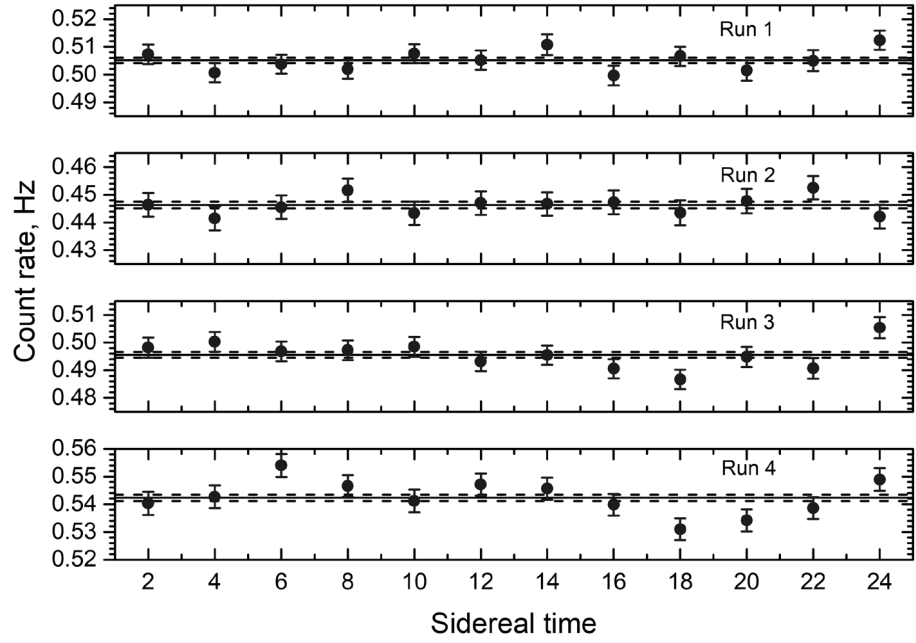

Fig. 4. Count rate of the detector with an Al-cathode (third series) in sidereal time (four runs).

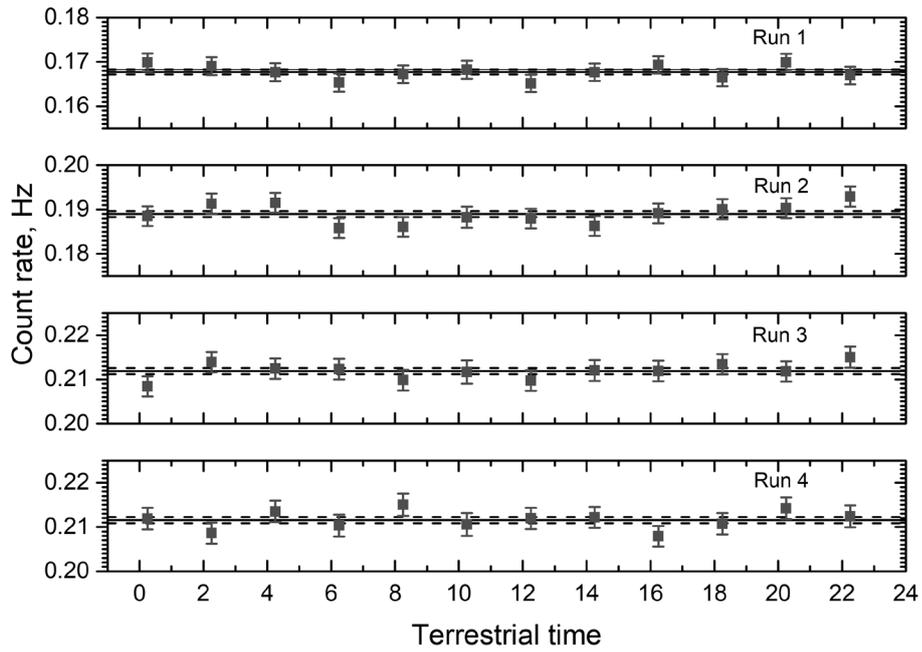

Fig. 5. Count rate of the detector with a Fe-cathode (third series) in solar time (four runs).

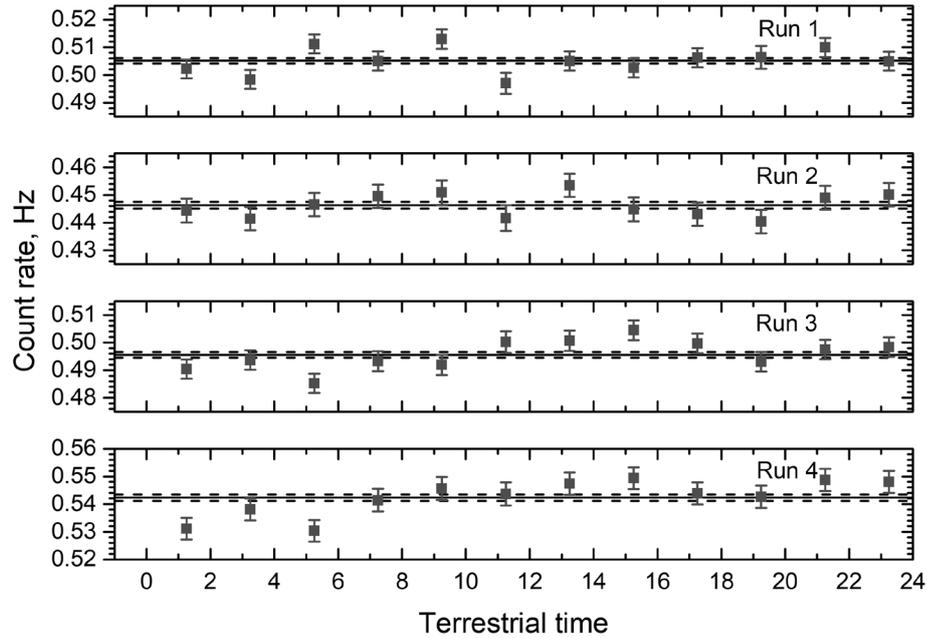

Fig. 6. Count rate of the detector with an Al-cathode (third series) in solar time (four runs).

Key result: On none of the detectors (Fe or Al), neither in sidereal nor in solar time, was a statistically significant 4-hour interval with an excess count rate above the average level found in each of the four runs, similar to those observed in the first two series [3, 4]. The same result is obtained if runs are grouped in pairs for the Fe detector and the Al detector for the first 120 days of measurements and for the last 120 days.

This result has several important implications:

1. Exclusion of instrumental/seasonal effects. The absence of a signal in the third series, conducted at a different time compared to the first series, argues against the previously observed excesses being caused by instrumental artifacts (e.g., temperature variations), terrestrial anthropogenic processes, or seasonal variation of Solar activity. The data from the third series show that the chosen criterion (the presence of an excess in the same 4-hour interval in all four runs) is indeed strong, thereby confirming the reliability of the chosen measurement methodology.

2. Confirmation of the non-persistence of the effect. The absence of a signal in the third series is consistent with the hypothesis of a "granulated" polarization structure of DM. The Sun, moving through the galaxy, could have left a region with a suitable orientation of the polarization vector (20°–50°) or entered a region with a different orientation that does not lead to the observed effect in this particular experimental setup. The difference in the timing of the excess between the first and second series also supports this idea.

3. No influence of cathode magnetic properties: Since the signal was absent on both active detectors (Fe and Al), based on these data, no conclusion can be drawn about the influence of the cathode's magnetic properties on the effect (if it exists).

The results of the three series can be interpreted as a possible effect occurring when the Sun passes through regions of dark matter with different orientations of the polarization vector.

Measurements in the PHELEX experiment continue using all three detectors with the aim of obtaining an evidence base.

## 4. CONCLUSION AND PROSPECTS

The conducted third measurement series in the PHELEX experiment with the expanded setup (two active detectors: Fe and Al cathodes) did not reveal the 4-hour excesses in the count rate of single electrons in sidereal time that were previously observed with high statistical significance in two separate series. The absence of the observed excess in the third series:

- Confirms the reliability of the methodology, ruling out persistent instrumental or seasonal effects as the cause of the previously observed excesses.
- Is consistent with the hypothesis of spatial "granulation" of dark matter according to the orientation of the dark photon polarization vector, leading to the episodic appearance of the predicted directional signal as the Sun moves through the galaxy.

The current sensitivity of the setup with one active detector allows recording the effect with high confidence if the Sun is in a "suitable" DM region (polarization vector ~20°–50° to the Earth's rotation axis) for at least ~200 days. To investigate domains of smaller spatio-temporal scale ("granulation spectrum" [6]), a significant increase in statistics is necessary.

A promising direction is the simultaneous operation of four identical detectors. This would allow:

1. Increasing the overall data acquisition rate by a factor of 4 while maintaining the data accumulation time for one "virtual run" at the level of ~50–60 days.
2. Achieving statistical significance comparable to the first series (~4–6σ) already for data accumulated during one run (~50–60 days).
3. Investigating the "granulation spectrum" of DM polarization with a time resolution of ~50–60 days, which corresponds to a spatial scale of ~$10^9$ km along the Sun's trajectory.

The continuation of measurements in the PHELEX experiment is a necessary step for confirming or refuting the hypothesis of a signal from dark photons and studying the possible domain structure of dark matter.

## ACKNOWLEDGEMENTS

The authors are grateful to the Ministry of Science and Higher Education for the substantial support within the framework of the "Equipment Renewal Program" under the state project "Science".

**REFERENCES**

1. W. J. R. Enzi, C. Krawczyk, D. Ballard, and T. Collett, *Mon. Not. Roy. Astron. Soc.* **540**, 247 (2025).

2. M. Cirelli, A. Strumia, and J. Zupan, arXiv: 2406.01705 [hep-ph].

3. A. V. Kopylov, I. V. Orekhov, and V. V. Petukhov, Phys. At. Nucl. **86**, 1296 (2023).

4. A. V. Kopylov, I. V. Orekhov, V. V. Petukhov, and A. E. Solomatin, Phys. At. Nucl. **87**, 810 (2024).

5. A. V. Kopylov, I. V. Orekhov, V. V. Petukhov, and A. E. Solomatin, Instrum. Exp.Tech. **68**, 7 (2025).

6. A. V. Kopylov, I. V. Orekhov, V. V. Petukhov, and A. E. Solomatin, arXiv: 2408.03620v2 [hep-ex].